\documentclass[runningheads]{svmult}

\usepackage{makeidx}   
\usepackage{graphicx}  
\usepackage{subeqnar}  
\usepackage{multicol}  
\usepackage{physprbb}  
\makeindex             

\begin{document}
\title*{Morphology and Redshifts of Extremely Red 
\protect\newline Galaxies in the GOODS/CDFS deep ISAAC field}
\toctitle{Morphology and Redshifts of Extremely Red 
\protect\newline Galaxies in the GOODS/CDFS deep ISAAC field}
%
%
\titlerunning{Morphology and Redshifts of Extremely Red Galaxies}
%
\author{Caputi K. I.
\and Dunlop J. S. 
\and McLure R. J.
\and Roche N. D.}
\authorrunning{Caputi K. I. et al.}
%
%
\institute{Institute for Astronomy, University of Edinburgh, 
Royal Observatory, 
\\
Edinburgh EH9 3HJ, Scotland.}

\maketitle              

\begin{abstract}
   We present the photometric redshift distribution of a sample of 198 Extremely Red Galaxies (ERGs) with $\rm K_s<22$ and $\rm (I_{775}-K_s)>3.92$ (Vega), selected by Roche et al. \cite{rch} in 50.4 arcmin$^2$ of the Chandra Deep Field South (CDFS). The sample has been obtained using ISAAC-VLT and ACS-HST GOODS public data. We also show the results of a morphological study of the 72 brightest ERGs in the z band ($\rm z<25$, AB). 
\end{abstract}

\section{Photometric Redshifts}

\begin{figure}[h]
\begin{center}
\includegraphics[width=1.0\textwidth]{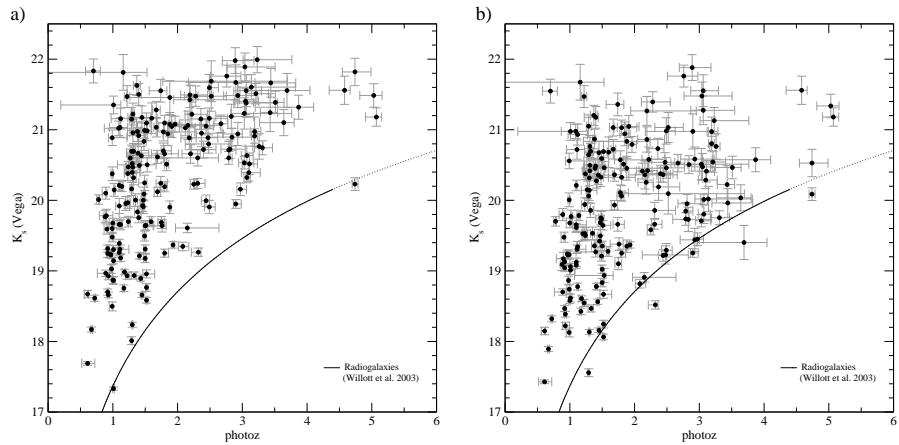}
\end{center}
\caption[]{Hubble diagram for the ERGs in Roche et al.'s sample. a) Original $\rm K_s$ magnitudes; b) dust-corrected $\rm K_s$ magnitudes. }
\label{Kz}
\end{figure}

 Figure \ref{Kz} shows the redshift distribution of  the ERGs in Roche et al.'s sample. The redshifts have been computed using the public code `hyperz' \cite{hyp} with photometry in seven bands: B, V, $\rm I_{775}$, z (V1.0 ACS-HST release); J,H and $\rm K_s$ (ISAAC-VLT). 

\noindent Our results show that the ERGs span a wide variety of luminosities in the Hubble diagram up to the high mass limit observationally defined by massive radio galaxies \cite{wil}.

\section{Morphology}
    
\begin{figure}[h]
\begin{center}
\includegraphics[width=.6\textwidth]{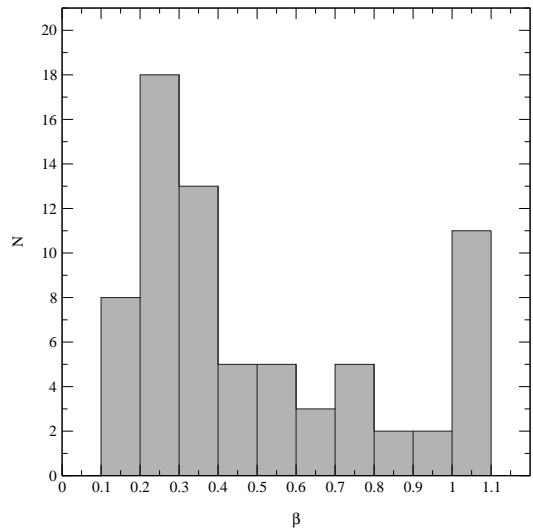}
\end{center}
\caption[]{Distribution of the $\rm \beta$ parameter for the 72 brightest ERGs in the z band.}
\label{morph}
\end{figure}

  The radial dependence of the surface brightness distribution of regular galaxies is given by $\rm I(r) \propto I_o \exp (-r/r_o)^{\beta}$, where $\rm I_o$ is  the surface brightness at a characteristic distance $\rm r_o$. Empirically, elliptical galaxies have $\beta \sim$ 0.25, while pure disks have $\beta \sim$ 1.  We determined the best-fit $\beta$ parameter of the 72 brightest ERGs in the $\rm z$-band, using a 2D-modelling code developed by McLure et al. \cite{mcl},  on the 5-epoch GOODS stacked images. The results are shown in fig. \ref{morph}.

\noindent We conclude that 61\% and 24\% of this subsample is composed of elliptical ($\rm 0< \beta<0.5$) and disky ($\rm 0.5<\beta<1.0$) galaxies, respectively.  The remaining 15\% are represented in the $\beta$ interval $[1.0,1.1]$ and are either irregular galaxies or a mix of disk and bulge.

%

\end{document}